\documentclass[conference, a4paper]{IEEEtran}

\usepackage{cite}
\usepackage{amsmath,amssymb,amsfonts}
\usepackage{algorithmic}
\usepackage{textcomp}
\usepackage{algorithm}
\usepackage{algorithmic}
\usepackage{mathrsfs}
\usepackage{booktabs}
\usepackage{graphicx}
\usepackage{subcaption}
\usepackage{longtable}
\usepackage{hyperref}
\usepackage{cleveref}
\usepackage{verbatim}
\usepackage{xcolor, soul}
\sethlcolor{yellow}

\input epsf

\begin{document}

\newtheorem{lemma}{Lemma}
\newtheorem{corol}{Corollary}
\newtheorem{theorem}{Theorem}
\newtheorem{proposition}{Proposition}
\newtheorem{definition}{Definition}

\title{A Smaller Subspace is All You Need: \\ Efficient ISAC Beamforming via Dimensionality Reduction}

\author{Ruiqi Liu\IEEEauthorrefmark{1}, Yuanshuo Gang\IEEEauthorrefmark{2}, Tianqi Mao\IEEEauthorrefmark{2}, and Qingqing Wu\IEEEauthorrefmark{4} \\   \IEEEauthorrefmark{1}Wireless and Computing Research Institute, ZTE Corporation, Beijing 100029, China \\ \IEEEauthorrefmark{2}State Key Laboratory of Environment Characteristics and Effects for Near-space, Beijing Institute of Technology, \\ Beijing 100081, China \\
\IEEEauthorrefmark{4}Department of Electronic Engineering, Shanghai Jiao Tong University, Shanghai 200240, China}

\maketitle

\begin{abstract}
Integrated sensing and communication (ISAC) has been recognized as a critical technology for the sixth generation (6G) wireless networks. Under monostatic sensing mode, achieving a balance between communication and sensing performance via optimal beamforming is a critical research problem. In multiple input multiple output (MIMO) systems, such beamforming problems are of substantial computational complexity due to the large number of antennas. Existing methods exploit exact subspaces of lower dimensions to speed up the beamforming process where the complexity is proportional to the number of communication users and sensing targets instead of antennas. In this paper, it is demonstrated that the beamforming problem can be solved in an even lower-dimensional subspace with only a small sensing performance loss. Numerical results compare the sensing performance and computational complexity of ISAC beamforming problem in the exact subspace and smaller subspaces.
\end{abstract}


\section{Introduction}\label{Sec:Intro}
Integrated sensing and communication (ISAC) has been identified as a promising technology for the next generation wireless communication, as well as a critical usage scenario of future communication systems \cite{10904090}. By reusing communication infrastructure for sensing, ISAC can reduce hardware and spectrum costs while improving spectral and energy efficiency compared to systems that rely on dedicated sensing resources \cite{mao2026movableantenna}. 

For ISAC systems, beamforming problem is critical for balancing performance between the communication and sensing functions. A common design principle is to satisfy all communication users for a certain signal-to-interference-plus-noise ratio (SINR) and in the meanwhile, to achieve a certain sensing performance. Among various sensing metrics, the Cram\'er-Rao bound (CRB) and the position error bound (PEB) are widely adopted in ISAC beamforming optimization problems.
For example, \cite{Liu2022CRB} investigated CRB minimization under communication SINR constraints, while \cite{Wang2023NFISAC} extended CRB based
ISAC design to near-field distance and angle estimation. Conventional optimization approaches for such beamforming problems include semidefinite relaxation (SDR) and interior-point method (IPM), which can incur high computational complexity when the number of transmit antennas is large. Thus, for multiple input multiple output (MIMO) ISAC systems, low-complexity beamforming is a critical topic.

To reduce the complexity, several studies have exploited the low-dimensional structures in ISAC beamforming problems. Authors of \cite{Lu2023Subspace} showed that the optimal ISAC transmit signal lies in the subspace spanned by the sensing steering directions and the communication channel subspace, which provides insightful directions for low-complexity ISAC beamforming. For multi-user MIMO ISAC on high frequencies, \cite{Wang2025LSBCD} exploited a low dimensional optimal precoder structure to develop an efficient block coordinate descent algorithm. Another recent study utilized the modular array structure to derive an optimal analog beamformer, thereby reducing hybrid beamforming to a lower dimensional digital optimization problem \cite{Meng2025Modular}.
More recently, \cite{Zhao2026LowDim} established a lossless low dimensional formulation for ISAC beamforming and developed reduced-dimensional interior point and augmented Lagrangian algorithms.
An exact subspace was constructed and a corresponding reduced-dimensional SDR was proposed for extended target CRB optimization in \cite{Zhao2026ET}, where numerical results demonstrated significant complexity reduction and runtime savings.
A similar effort was reported by \cite{Fang2026Structure}, where authors constructed a subspace to decrease the dimension of the optimization problem in a far-field ISAC system. The dimension of the problem is reduced from the number of transmit antenna or number of transmitted symbols, which are normally large, to the number of communication users plus three times the number of sensing targets. Simulations demonstrated that the proposed low-dimensional beamforming design achieves the same communication and sensing tradeoff frontier while decreasing the runtime.
In another direction, \cite{Lee2026SDISAC} employed the communication channel null space to decouple sensing and communication waveform design. These studies focused on constructing an exact and lossless subspace in which the search for optimal beamformer could be done in a faster way. 
To the best of the authors' knowledge, the feasibility and performance of pruning an exact and lossless subspace to attain even smaller beamforming dimensions has not been studied.

Motivated by this gap, this paper moves beyond conventional exact subspace construction and investigates whether the beamforming search space can be compressed further with limited sensing performance degradation. The objective is to reduce the optimization dimension in a solver-agnostic manner, so that subspace compression can be combined with different beamforming optimization methods.
To this end, two singular value decomposition (SVD) based pruning methods are studied: communication preserving SVD that retains the entire communication subspace and global SVD that jointly selects dominant communication and sensing directions. Both are incorporated into a reduced-dimensional optimization problem which is solved using SDR and IPM. By evaluating the PEB, communication feasibility and computational runtime, the feasibility and limitations of dimensionality reduction beyond the exact communication-sensing subspace is revealed and analyzed.

The rest of this paper is organized as follows. The system model and ISAC beamforming problem are described in Section \ref{sec:system}. The exact subspace construction and further dimensionality reduction methods are proposed in Section \ref{sec:subspace}. In Section \ref{Sec:simulation}, simulation results are provided to demonstrate the feasibility and performance aspects of the proposed dimensionality reduction operation. Finally, Section \ref{sec:conclusion} concludes the paper.

\textit{Notations:} 
$(\cdot)^T$, $(\cdot)^H$, $(\cdot)^*$ and $\|\cdot\|_F$ denote transpose, conjugate transpose, complex conjugate and Frobenius norm, respectively. $\mathbb C$ and $\mathbb R$ are the complex and real fields.
$\mathbb E$ denotes expectation. $\operatorname{range}(\cdot)$, $\operatorname{rank}(\cdot)$ and $\operatorname{tr}(\cdot)$ denote column span, rank and trace of a matrix. $\mathbf I_K$ is the $K\times K$ identity matrix. $j=\sqrt{-1}$ denotes the imaginary unit.

\section{System Model and Problem Formulation}\label{sec:system}
In this paper, a monostatic ISAC system is considered where a base station (BS) with a uniform planar array (UPA) of $N_t=N_xN_y$ antennas serves $K$ single-antenna communication users and localizes one point target. The array contains $N_x$ and $N_y$ elements along the $x$- and $y$-axes, respectively, with half-wavelength spacing $\lambda/2$, where $\lambda$ is the carrier wavelength. With the BS located at the origin, the position of antenna $(n_x,n_y)$ can be expressed as
\begin{equation}
\mathbf p_{n_x,n_y}=\frac{\lambda}{2}
\begin{bmatrix}
n_x-(N_x+1)/2, & n_y-(N_y+1)/2, & 0
\end{bmatrix}^T,
\end{equation}
where $n_x\in\{1,\ldots,N_x\}$ and $n_y\in\{1,\ldots,N_y\}$. For the sake of notational simplicity, $(n_x,n_y)$ is mapped to a scalar index $n=(n_y-1)N_x+n_x, n\in\{1,\ldots,N_t\}.$

To describe the locations of both communication users and the sensing target, consider an arbitrary spatial point with spherical coordinates $\boldsymbol\eta=[r,\theta,\phi]^T$ relative to the center of the BS antenna array, where $r$ is the range, $\theta$ is the polar angle measured from the positive $z$-axis and $\phi$ is the azimuth. Equivalently, its Cartesian position is
\begin{align}
\mathbf q(\boldsymbol\eta)
&=r[\sin\theta\cos\phi,\,\sin\theta\sin\phi,\,\cos\theta]^T,
\end{align}
and the propagation distance from this point to the $n$-th BS antenna element located at $\mathbf p_n$ is
\begin{align}
d_n(\boldsymbol\eta)
&=\|\mathbf q(\boldsymbol\eta)-\mathbf p_n\|_2,
\end{align}
respectively.
The array response is formulated using exact spherical wave phase differences for both near- and far-field points as follows:
\begin{equation}
[\mathbf a(\boldsymbol\eta)]_n
=\exp\!(-j\frac{2\pi}{\lambda}[d_n(\boldsymbol\eta)-r]).
\end{equation}
Following the commonly used uniform-amplitude approximation, the large-scale propagation gain from a given spatial point to all BS antenna elements is assumed to be identical, while the element-dependent propagation phases are calculated independently. In the far field, this response naturally reduces to the conventional plane-wave steering model.

\subsection{Communication and Sensing Signal Models}
In this paper, the BS serves $K$ communication users with the beamforming matrix
$\mathbf W=[\mathbf w_1,\ldots,\mathbf w_K]\in\mathbb C^{N_t\times K}$,
where $\mathbf w_k$ is the beamformer for user $k\in\{1,\ldots,K\}$. At snapshot $\ell\in\{1,\ldots,L\}$, the transmitted signal from the BS is
\begin{equation}
\mathbf x[\ell]=\sum_{k=1}^K\mathbf w_ks_k[\ell]
=\mathbf W\mathbf s[\ell],
\end{equation}
where $s_k[\ell]$ denotes the symbol for user $k$ and
$\mathbf s[\ell]=[s_1[\ell],\ldots,s_K[\ell]]^T$.
The symbols intended for different users are assumed to be mutually uncorrelated and normalized to unit power such that
$\mathbb E \{\mathbf s[\ell]\mathbf s^H[\ell]\} = \mathbf I_K$.
Consequently, the transmit covariance matrix is
\begin{equation} \label{eq:R_x_W_W}
\mathbf R_x \triangleq \mathbb E\{\mathbf x[\ell]\mathbf x^H[\ell]\} = \mathbf W\mathbf W^H.
\end{equation}

Denote the position of user $k$ as $\boldsymbol\eta_k=[r_k,\theta_k,\phi_k]^T$. 
The communication channel is defined as $\mathbf h_k=\alpha_k^*\mathbf a^*(\boldsymbol\eta_k)$, where $\alpha_k$ denotes the complex path gain.
Then, the received signal and SINR are expressed as
\begin{align}
y_k[\ell]&=\mathbf h_k^H\mathbf x[\ell]+n_k[\ell], \\
\gamma_k(\mathbf W)&=
\frac{|\mathbf h_k^H\mathbf w_k|^2}{\sum_{i\ne k}|\mathbf h_k^H\mathbf w_i|^2+\sigma_k^2},\label{eq:communication_sinr}
\end{align}
where $n_k[\ell]\sim\mathcal{CN}(0,\sigma_k^2)$ is complex Gaussian noise with variance $\sigma_k^2$ and $i\in\{1,\ldots,K\}$ indexes interfering users.

For a sensing target with position $\boldsymbol\eta_s=[r_s,\theta_s,\phi_s]^T$, define $\mathbf g_s=\mathbf a(\boldsymbol\eta_s)$. Under monostatic sensing, the two-way response is $\mathbf G_s=\mathbf g_s\mathbf g_s^T\in\mathbb C^{N_t\times N_t}$. The received echo at the BS is
\begin{equation}
\mathbf y_s[\ell]=\beta_s\mathbf G_s\mathbf x[\ell]+\mathbf n_s[\ell],
\end{equation}
where $\beta_s$ represents the unknown complex target reflection coefficient and $\mathbf n_s[\ell]\sim\mathcal{CN}(\mathbf 0,\sigma_s^2\mathbf I_{N_t})$ represents receiver noise of variance $\sigma_s^2$ per antenna.

\subsection{Sensing Metric Characterization and Problem Formulation}
In the ISAC system considered in this paper, the target position is estimated with the reflection coefficient $\beta_s$ unknown. Let $b_R=\Re\{\beta_s\}$ and $b_I=\Im\{\beta_s\}$ denote its real and imaginary parts, respectively. Define the real parameter vector $\boldsymbol\xi=[\boldsymbol\eta_s^T,b_R,b_I]^T\in\mathbb R^5$. 
Let $\boldsymbol\mu[\ell;\boldsymbol\xi] = \beta_s\mathbf G_s\mathbf W\mathbf s[\ell]$ denote the noiseless sensing observation at snapshot $\ell$. For $L$ statistically independent snapshots, the Fisher information matrix (FIM) can be written element-wise as
\begin{equation}
[\mathbf J]_{a,b}
= \frac{2}{\sigma_s^2}
\sum_{\ell=1}^{L}
\mathbb E [ \Re \{ (\frac{\partial\boldsymbol\mu[\ell]}{\partial\xi_a})^H
\frac{\partial\boldsymbol\mu[\ell]}{\partial\xi_b}
\}] ,
\end{equation}
where $a,b\in\{1,\ldots,5\}$ index the entries of $\boldsymbol\xi$. Note that $\mathbf J\in\mathbb R^{5\times5}$. For $u\in\{1,2,3\}$, let $\eta_{s,u}$ denote the $u$-th component of $\boldsymbol\eta_s$ and define
\begin{equation} \label{eq:G_s_u}
\mathbf G_{s,u}=\partial\mathbf G_s/\partial\eta_{s,u}.
\end{equation}
For $u,v\in\{1,2,3\}$, define $\Omega_{u,v}=\operatorname{tr}(\mathbf G_{s,v}\mathbf R_x\mathbf G_{s,u}^H)$, $\zeta_u=\operatorname{tr}(\mathbf G_s\mathbf R_x\mathbf G_{s,u}^H)$ and $\nu=\operatorname{tr}(\mathbf G_s\mathbf R_x\mathbf G_s^H)$ where $\mathbf R_x = \mathbf W\mathbf W^H$ according to \eqref{eq:R_x_W_W}.

Then, the FIM $\mathbf J$ can be partitioned into $3\times3$, $3\times2$, and $2\times2$ blocks as
\begin{equation}\label{eq:fim_partition}
\mathbf J=\begin{bmatrix}
\mathbf J_{\eta\eta}&\mathbf J_{\eta\beta}\\
\mathbf J_{\beta\eta}&\mathbf J_{\beta\beta}
\end{bmatrix},
\end{equation}
where 
$[\mathbf J_{\eta\eta}]_{u,v} =\frac{2L|\beta_s|^2}{\sigma_s^2}\Re\{\Omega_{u,v}\}$, $[\mathbf J_{\eta\beta}]_{u,1} =\frac{2L}{\sigma_s^2}\Re\{\beta_s^*\zeta_u\}$, $[\mathbf J_{\eta\beta}]_{u,2}=-\frac{2L}{\sigma_s^2}\Im\{\beta_s^*\zeta_u\}$, $\mathbf J_{\beta\beta}=\frac{2L\nu}{\sigma_s^2}\mathbf I_2$ and $\mathbf J_{\beta\eta}=\mathbf J_{\eta\beta}^T$.
Specifically, $u,v\in\{1,2,3\}$ index the three target coordinates $\eta_{s,1}=r_s, \eta_{s,2}=\theta_s, \eta_{s,3}=\phi_s.$
Hence, $\mathbf J_{\eta\eta}\in\mathbb R^{3\times3}$ characterizes the Fisher information and coupling among range, polar angle, and azimuth. The first and second columns of $\mathbf J_{\eta\beta}\in\mathbb R^{3\times2}$ characterize their coupling with the nuisance parameters $b_R=\Re\{\beta_s\}$ and $b_I=\Im\{\beta_s\}$, respectively.

Eliminating the nuisance parameters through the Schur complement gives the equivalent FIM (EFIM) with respect to a given beamforming matrix as
\begin{equation}
\mathbf J_{\rm e}(\mathbf W)=\mathbf J_{\eta\eta}-\mathbf J_{\eta\beta}
\mathbf J_{\beta\beta}^{-1}\mathbf J_{\beta\eta}.
\end{equation}
The target coordinate CRB is $\mathbf C_\eta(\mathbf W)=\mathbf J_{\rm e}^{-1}(\mathbf W)$ when the EFIM is nonsingular. Define $\mathbf D_\eta=\partial\mathbf q(\boldsymbol\eta_s)/\partial\boldsymbol\eta_s$ and $\boldsymbol\Lambda_\eta=\mathbf D_\eta^T\mathbf D_\eta=\operatorname{diag}(1,r_s^2,r_s^2\sin^2\theta_s)$. The squared position error bound (SPEB) is
\begin{equation}
\operatorname{SPEB}(\mathbf W)=\operatorname{tr}(\boldsymbol\Lambda_\eta\mathbf J_{\rm e}^{-1}(\mathbf W)) = \operatorname{PEB}^2(\mathbf W).
\end{equation}
Since PEB and SPEB have the same minimizers, the beamforming problem is formulated as
\begin{subequations}
\begin{align}
\text{(P1)}:\quad\min_{\mathbf W}\quad&\operatorname{SPEB}(\mathbf W) \\
\mathrm{s.t.}\quad&\gamma_k(\mathbf W)\ge\Gamma_k,\ \forall k, \\
&\|\mathbf W\|_F^2\le P_{\max},
\end{align}
\end{subequations}
where $\Gamma_k$ is the minimum SINR required for user $k$ and $P_{\max}$ is the maximum transmit power of the BS. The FIM and sensing subspace are evaluated at a nominal target position. 

\section{Subspace Construction for ISAC Beamforming}
\label{sec:subspace}

\subsection{Exact Communication-Sensing Subspace}
Low-dimensional subspaces for ISAC beamforming have been established for several problems \cite{Zhao2026LowDim,Zhao2026ET,Fang2026Structure}. For the ISAC model in Section \ref{sec:system}, an exact subspace is constructed as follows. Note that an exact subspace means that a high-dimensional dataset can be perfectly embedded into a lower-dimensional linear space without losing any information, and thus, restricting the beamforming search to this subspace does not change the optimal value of (P1).

Define $\mathbf H=[\mathbf h_1,\ldots,\mathbf h_K]\in\mathbb C^{N_t\times K}$ as the communication channel matrix. For each target coordinate $\eta_{s,u}$, $u\in\{1,2,3\}$, define
$\mathbf g_{s,u}=\frac{\partial\mathbf g_s}{\partial\eta_{s,u}}$ and with \eqref{eq:G_s_u}, $\mathbf G_{s,u}=\mathbf g_{s,u}\mathbf g_s^T+\mathbf g_s\mathbf g_{s,u}^T$.

From $\mathbf G_s=\mathbf g_s\mathbf g_s^T$ and $\mathbf G_{s,u}=\mathbf g_{s,u}\mathbf g_s^T+ \mathbf g_s\mathbf g_{s,u}^T$, the dependence of the sensing response and its derivatives on $\mathbf w_k$ is completely characterized by the scalar projections $\mathbf g_s^T\mathbf w_k$ and $\mathbf g_{s,u}^T\mathbf w_k$. Hence, preserving the subspace spanned by $\mathbf g_s^*$, $\mathbf g_{s,1}^*$, $\mathbf g_{s,2}^*$ and $\mathbf g_{s,3}^*$ is sufficient to preserve all elements in  \eqref{eq:fim_partition}.
Thus, these four vectors are referred to as the sensing subspace generators and are collected into
\begin{equation}
\mathbf S_s
= [\mathbf g_s^*,\mathbf g_{s,1}^*,
\mathbf g_{s,2}^*,\mathbf g_{s,3}^*]
\in\mathbb C^{N_t\times4}.
\end{equation}

Combining the communication and sensing directions, define the composite generator matrix
$\mathbf A_{\rm ex} = [\mathbf H,\mathbf S_s]$.
Its column space defines the exact communication-sensing subspace
\begin{equation} \label{eq:exact_subspace}
\mathcal S_{\rm ex} = \operatorname{range}(\mathbf A_{\rm ex}).
\end{equation}
Note that the dimension of the exact subspace is $d_{\rm ex} = \operatorname{rank}(\mathbf A_{\rm ex}) \le \min\{N_t,K+4\}.$ For generic user and target geometries with linearly independent generators and $N_t\ge K+4$, $d_{\rm ex} = K+4$.
Let $\mathbf U_{\rm ex}$ be an orthonormal basis of
$\mathcal S_{\rm ex}$.
For a particular beamformer $\mathbf w_k$, it can be decomposed as
\begin{equation}
\mathbf w_k=\mathbf U_{\rm ex}\mathbf v_k+\mathbf z_k,
\end{equation}
where $\mathbf v_k\in\mathbb C^{d_{\rm ex}}$ is the reduced beamformer and $\mathbf z_k$ is its orthogonal complement such that $\mathbf U_{\rm ex}^H\mathbf z_k=\mathbf 0$. This implies that $\mathbf h_i^H\mathbf z_k=0$, $\mathbf g_s^T\mathbf z_k=0$ and $\mathbf g_{s,u}^T\mathbf z_k=0$. 
$\mathbf h_i^H\mathbf z_k=0$ effectively means that $\mathbf z_k$ affects neither the desired signal nor the multi-user interference. For sensing, it can be seen that
\begin{equation}
\mathbf G_s\mathbf z_k =\mathbf g_s(\mathbf g_s^T\mathbf z_k)=\mathbf 0, \end{equation}
\begin{equation}
\mathbf G_{s,u}\mathbf z_k =\mathbf g_{s,u}(\mathbf g_s^T\mathbf z_k)
+\mathbf g_s(\mathbf g_{s,u}^T\mathbf z_k)
=\mathbf 0.
\end{equation}
Hence, replacing $\mathbf w_k$ by $\mathbf U_{\rm ex}\mathbf v_k$ preserves communication SINRs and sensing FIM elements. Meanwhile, the orthogonal decomposition gives
\begin{equation}
\|\mathbf w_k\|_2^2 = \|\mathbf v_k\|_2^2+\|\mathbf z_k\|_2^2,
\end{equation}
so the transmit power cannot increase after the projection. Consequently, an optimal solution exists in the exact subspace $\mathcal S_{\rm ex}$ and admits
\begin{equation}
\mathbf W=\mathbf U_{\rm ex}\mathbf V,\qquad
\mathbf V\in\mathbb C^{d_{\rm ex}\times K}.
\end{equation}

\subsection{Further Dimensionality Reduction via Subspace Pruning}
Although restricting the beamforming search to $\mathcal S_{\rm ex}$ is lossless, this subspace may contain numerically weak or correlated directions. This motivates the operation of further dimensionality reduction by pruning the exact subspace, which already has fewer dimensions compared to the full space of the original problem. For this purpose, two SVD based approaches are proposed to achieve a dimension lower than $d_{\rm ex}$. These two methods are chosen to be as general as possible.

\subsubsection{Communication Preserving SVD}
A conservative strategy for further reducing $\mathcal S_{\rm ex}$ is to preserve the entire communication subspace and prune only the sensing components outside it. Since $\operatorname{range}(\mathbf H)$ is fully retained, $\mathbf h_k^H\mathbf w_i$ in the SINR expression in \eqref{eq:communication_sinr} is preserved for all $k$ and $i$ under orthogonal projection onto the SVD subspace. Hence, the pruning operation itself does not reduce the communication SINRs.

Denote $r_c=\operatorname{rank}(\mathbf H)$ and let
$\mathbf U_c\in\mathbb C^{N_t\times r_c}$ be an orthonormal basis of $\operatorname{range}(\mathbf H)$. The residual sensing generators can be written as
\begin{equation}\label{eq:sensing_residual}
\mathbf S_\perp=(\mathbf I_{N_t}-\mathbf U_c\mathbf U_c^H)\mathbf S_s
=[\mathbf s_{\perp,1},\ldots,\mathbf s_{\perp,4}].
\end{equation}
In other words, $\mathbf S_\perp$ contains the sensing related components that lie outside the communication subspace. Essentially, $\mathbf S_\perp$ characterizes the additional spatial degrees of freedom required by sensing beyond those already provided for communication.

The four residual generators may have different magnitudes. Directly applying SVD may therefore cause the decomposition to be dominated by their raw numerical magnitudes rather than their directional correlations. To remove this scale dependence, every nonzero residual column is normalized to unit norm as
\begin{equation}
\widetilde{\mathbf s}_{\perp,m}=
\frac{\mathbf s_{\perp,m}}{\|\mathbf s_{\perp,m}\|_2} .
\end{equation}
Combining all four normalized sensing residual columns gives $\widetilde{\mathbf S}_\perp
=[\widetilde{\mathbf s}_{\perp,1},\ldots,
\widetilde{\mathbf s}_{\perp,4}]$. SVD operation is taken upon $\widetilde{\mathbf S}_\perp$ as
\begin{equation}
\widetilde{\mathbf S}_\perp=\mathbf U_s\mathbf\Sigma_s\mathbf V_s^H,
\end{equation}
where $\mathbf U_s$ and $\mathbf V_s$ contain the left and right singular vectors, respectively, and $\mathbf\Sigma_s$ contains descending singular values. With the SVD, the basis of communication preserving subspace is
\begin{equation}\label{eq:comm_preserving_basis}
\mathbf U_{\rm CP}(d)=[\mathbf U_c,\mathbf U_s(:,1:d-r_c)],
\quad r_c\le d\le d_{\rm ex}.
\end{equation}
The notation $\mathbf U_s(:,1:q)$ denotes the first $q$ columns of $\mathbf U_s$. This essentially means that the $(d-r_c)$ left singular vectors in $\mathbf U_s$ associated with largest singular values are selected to combine with $\mathbf U_c$ to form the $d$-dimensional subspace.
Since $\operatorname{range}(\mathbf H)\subseteq\operatorname{range}(\mathbf U_{\rm CP})$, the SINR of any feasible beamformer is preserved by projection onto the SVD space with non-increasing power. 

\subsubsection{Global SVD}
The second pruning method is even more straightforward. It simply compresses communication and sensing directions jointly by taking a SVD over the exact subspace $\mathcal S_{\rm ex}$ defined in \eqref{eq:exact_subspace}. 
Express $\mathbf A_{\rm ex}$ as $\mathbf A_{\rm ex}=[\mathbf a_1,\ldots,\mathbf a_{K+4}]$ and normalize its nonzero columns as $\widetilde{\mathbf a}_m=\mathbf a_m/\|\mathbf a_m\|_2$, for $m\in\{1,\ldots,K+4\}$. With $\widetilde{\mathbf A}_{\rm ex}$ denoting the normalized matrix comprising $\widetilde{\mathbf a}_m$, perform SVD as
\begin{equation}
\widetilde{\mathbf A}_{\rm ex}=\mathbf U_g\mathbf\Sigma_g\mathbf V_g^H,
\end{equation}
where $\mathbf U_g$, $\mathbf\Sigma_g$, and $\mathbf V_g$ are the corresponding left singular vectors, singular values, and right singular vectors. To prune to dimension $d$, simply retain the $d$ left singular vectors with largest singular values as
\begin{equation}\label{eq:global_basis}
\mathbf U_{\rm G}(d)=\mathbf U_g(:,1:d),\qquad 1\le d\le d_{\rm ex}.
\end{equation}
Unlike communication preserving SVD, global SVD might discard communication directions. It may therefore lead to a more compact representation of communication and sensing directions, but it may fail the SINR constraints for some cases.

\subsubsection{Reduced-Dimensional Beamforming}
Let $\mathbf U_d\in\mathbb C^{N_t\times d}$ denote the basis used. Depending on the pruning method employed, $\mathbf U_d$ can be \eqref{eq:comm_preserving_basis} or \eqref{eq:global_basis}. 
Setting $\mathbf W=\mathbf U_d\mathbf V$ yields
\begin{subequations}\label{eq:p2}
\begin{align}
\text{(P2)}:\quad\min_{\mathbf V}\quad&\operatorname{SPEB}(\mathbf U_d\mathbf V) \\
\mathrm{s.t.}\quad&\gamma_k(\mathbf U_d\mathbf V)\ge\Gamma_k,\ \forall k, \\
&\|\mathbf V\|_F^2\le P_{\max}.
\end{align}
\end{subequations}

To solve (P2), optimization methods such as IPM and SDR based semi-definite programming (SDP) are used, which take iterations to minimize the objective functions. For brevity, they are denoted by ``IPM'' and ``SDR'', respectively, in the numerical results.
For initialization, let $\overline{\mathbf H}=\mathbf U_d^H\mathbf H \in\mathbb C^{d\times K}$. A communication zero forcing (ZF) initializer can be written as
\begin{equation}\label{eq:zf_initialization}
\mathbf V_{\rm ZF}=\overline{\mathbf H}
(\overline{\mathbf H}^H\overline{\mathbf H})^{-1}
\operatorname{diag}(\sqrt{\Gamma_1\sigma_1^2},\ldots,
\sqrt{\Gamma_K\sigma_K^2}).
\end{equation}
Note that other advanced initializers are feasible while communication ZF is selected to represent the most general case possible.


The computational complexity analysis for the proposed reduced-dimensional beamforming can be found in Appendix \ref{app:complexity_analysis}.

\section{Numerical Results}\label{Sec:simulation}
In this Section, simulation results are provided to demonstrate the feasibility of further dimensionality reduction beyond exact subspace as well as the performance and runtime analysis. 
Since the benefits of lossless reduction from the full beamforming space to an exact subspace has been well demonstrated in prior studies \cite{Zhao2026ET,Zhao2026LowDim,Fang2026Structure},  the exact subspace is used as the baseline and simulations in this Section focus on further reducing its dimension. 
Two different subspace pruning methods are tested with two optimization solvers. 

In the simulations, the BS is equipped with 64 antennas arranged in an $8 \times8$ UPA. There are 3 communication users and one sensing target in each simulation. The center frequency is 28 GHz, resulting in a Rayleigh distance of 0.525 m. In each Monte Carlo test, the ranges of all users and the target are independently drawn from a log-uniform distribution over $[0.1,10]$ m. Their polar angles and azimuth angles are independently and uniformly generated from $[15^\circ,70^\circ]$
and $[‑80^\circ,80^\circ]$, respectively. The exact
spherical‑wave phase model is used for every user and target. The SINR requirement is $10$ dB for each user, the maximum BS transmit power is $P_{\max}=1$ W, and the communication noise variance is $\sigma_k^2=10^{-3}$ W. The sensing FIM is evaluated using $L=128$ snapshots with sensing noise variance $\sigma_s^2=1$ W. 10 beamforming iterations are mandated for IPM and SDR to observe converged PEB values. All reported results are obtained from 50 Monte Carlo tests. 
The simulations are conducted in Matlab 2026b on a laptop with a single core 1.7 GHz CPU. IPM is implemented using fmincon solver in Matlab optimization toolbox. The SDR based approach employs sequential rank-one penalization and recovery, while its convex SDP subproblems are solved by MOSEK~\cite{mosek}.

\begin{figure}[t]
    \centering
    \includegraphics[width=0.48\textwidth]{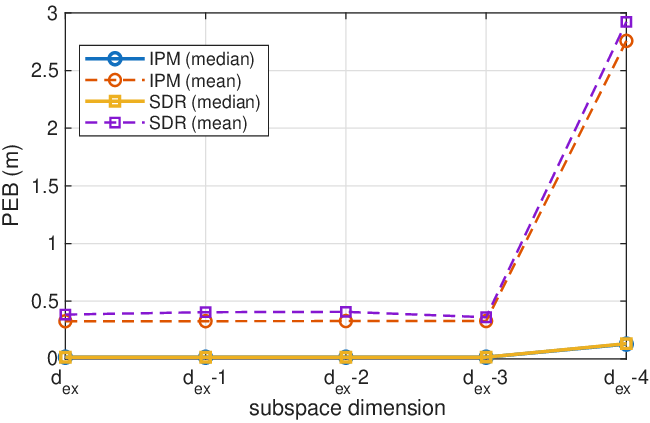}
    \caption{Mean and median PEBs when using communication preserving SVD to prune the exact subspace.}
    \label{fig:comm_svd_PEB}
    \vspace{-2 mm}
\end{figure}

\begin{figure}[t]
    \centering
\includegraphics[width=0.48\textwidth]{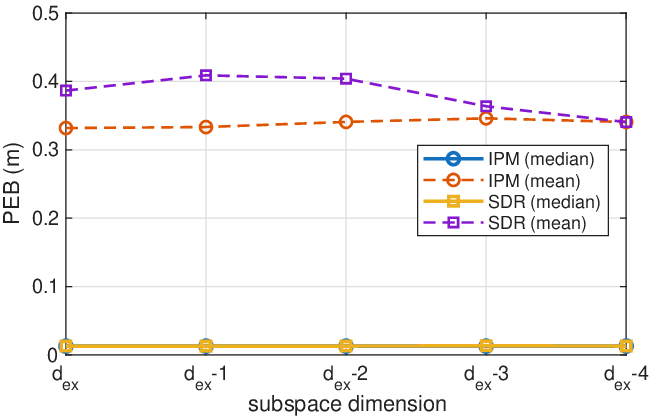}
    \caption{Mean and median PEBs when using global SVD to prune the exact subspace.}
    \label{fig:global_svd_PEB}
    \vspace{-3 mm}
\end{figure}

The mean and median PEBs obtained from both IPM and SDR using different subspace pruning methods are given in Fig. \ref{fig:comm_svd_PEB} and Fig. \ref{fig:global_svd_PEB}. It can be seen that under both pruning methods, when the subspace dimension is decreased from $d_{ex}$ to $(d_{ex} -3)$, the mean and median PEBs obtained by both solvers maintain a same level and the degradations are limited. Median PEBs obtained by both solvers are approximately 0.014 m under communication preserving SVD and 0.013 m under global SVD till dimension $(d_{ex} -3)$. 
For SDR solver, in some cases, the mean PEBs obtained from a lower dimensional subspace are even lower than those from a higher dimensional subspace. 
This observation should not be interpreted as performance advantage of the smaller subspace, since every truncated subspace is contained in the exact subspace. 
One potential reason behind such improvement is that after pruning, the numerical condition for SDR solver is improved since a lower dimension serves as regularization for the optimization problem.

For communication preserving pruning, it can be seen from Fig. \ref{fig:comm_svd_PEB} that the mean PEBs achieved for SDR and IPM when $d = d_{ex}-4$ are over 2.75 m. The median PEBs, on the other hand, are around 0.13 m when $d = d_{ex}-4$. This indicates that a few outlier cases where the optimal beamformer lies outside the $(d_{ex}-4)$-dimensional space and thus cannot be obtained contribute significantly to the mean PEB.

The results also show that global SVD, a more straightforward way to prune without delicate considerations, can maintain the mean and median PEB at a same level even when pruning 4 dimensions. For the considered $K=3$ setting, $d_{\rm ex}\le K+4=7$. The composite generator matrix has numerical rank seven in all 50 simulated drops, and hence $d_{\rm ex}=7$ throughout the reported results. The most aggressive pruning therefore reduces the search dimension from 7 to 3.
However, global SVD pruning cannot guarantee SINR requirements met for all communication users and may cause the beamformer to fail at its initialization when the dimension is below $(d_{ex} -3)$, which happens for both IPM and SDR. According to the simulations conducted for this paper, 47 out of 50 tests under global SVD pruning succeed. Note that to ensure fair comparison between two pruning methods, although communication preserving SVD succeeds in all 50 cases, the same 47 cases where global SVD pruning succeeds are selected to produce Fig. \ref{fig:comm_svd_PEB}.

\begin{figure}[t]
    \centering
    \includegraphics[width=0.48\textwidth]{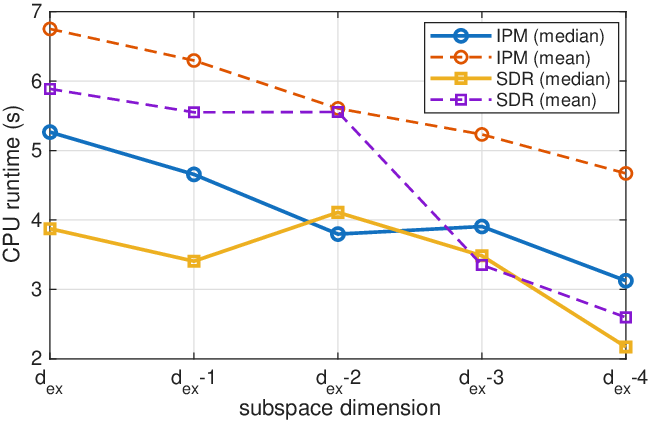}
    \caption{CPU runtime comparison under global SVD pruning for the successful cases.}
    \label{fig:CPU_time}
    \vspace{-2 mm}
\end{figure}

\begin{figure}[t]
    \centering
    \includegraphics[width=0.48\textwidth]{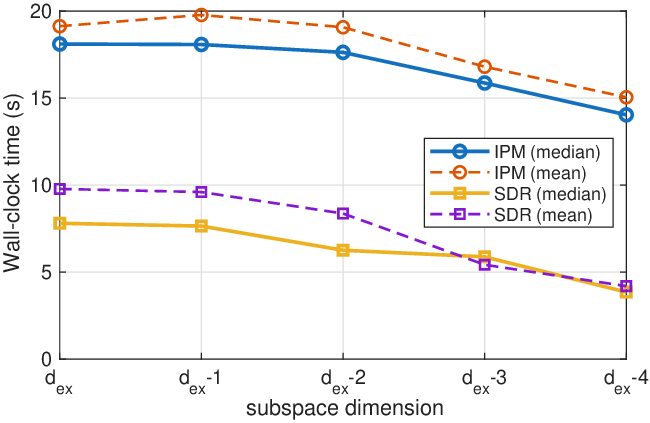}
    \caption{Wall-clock time comparison under global SVD pruning for the successful cases.}
    \label{fig:wall_clock}
    \vspace{-3 mm}
\end{figure}

Since global SVD pruning can be extended even to $(d_{ex} -4)$ dimensional case with a satisfying PEB, it is selected to demonstrate for runtime analysis. It can be seen from Fig. \ref{fig:CPU_time} that the central processing unit (CPU) time for each simulation drop consisting of 10 optimization iterations significantly decreases as the dimensionality of the problem decreases. In an average sense, the CPU time for IPM decreases from 6.75 seconds to 4.67 seconds and the CPU time for SDR decreases from 5.88 seconds to 2.60 seconds when the dimension is pruned from $d_{ex}$ to $(d_{ex}-4)$. In practical scenarios where timely decisions are to be made and computational resources are limited, such reductions are valuable.
A similar decreasing trend can be observed for the overall execution time for each simulation drop as depicted in Fig. \ref{fig:wall_clock}. 

\section{Conclusion}\label{sec:conclusion}
This paper investigates further dimensionality reduction for ISAC beamforming beyond an exact and lossless subspace. Two SVD based pruning methods are proposed, namely, communication preserving SVD, which retains the complete communication subspace, and global SVD, which selects dominant directions jointly from communication and sensing generators. The resulting reduced-dimensional problems are solved using both SDR and IPM. Numerical results show that substantial additional dimensionality reduction can be achieved with a small PEB loss while providing noticeable runtime savings. The results suggest that the practical dimension required for ISAC beamforming can be smaller than that of a lossless representation. The benefits of complexity reduction can be harnessed by at least two subspace construction methods and two commonly used beamforming algorithms under randomized settings, which hints that the idea of dimensionality reduction with limited performance loss has the potential of being generally applicable to many cases. 

\begin{appendices}
\section{Computational Complexity Analysis}\label{app:complexity_analysis}
Constructing an orthonormal basis of $\mathbf A_{\rm ex}\in\mathbb C^{N_t\times(K+4)}$ through QR decomposition or SVD requires $\mathcal O(N_t(K+4)^2)$ calculations. For communication preserving SVD, constructing $\mathbf U_c$, projecting the four sensing generators and performing the SVD of the $N_t\times4$ residual matrix require $\mathcal O(N_tK^2+4N_tK+16N_t)$ calculations. For global SVD, normalizing the columns of $\mathbf A_{\rm ex}$ along with its SVD require totally $\mathcal O(N_t(K+4)^2)$ calculations. These costs are incurred only once before beamforming optimization.

The main complexity reduction therefore results from replacing the $N_t$- or $d_{\rm ex}$-dimensional beamforming variables by their $d$-dimensional counterparts.
For IPM, each iteration mainly requires solving a Newton system whose size grows with the number of beamforming variables. Since $\mathbf V\in\mathbb C^{d\times K}$ contains $2dK$ real variables, the computational complexity is $\mathcal O((dK)^3)$.
For the SDR based approach, the $K$ beamforming vectors are lifted to $K$ positive semi-definite matrices of size $d\times d$. Hence, the lifted formulation contains $\mathcal O(Kd^2)$ scalar optimization variables. When the resulting SDP is solved by an interior point solver, the corresponding complexity is approximately $\mathcal O((Kd^2)^3)$.
These expressions are just intended to characterize the dependence on $d$, rather than the exact runtime of algorithms. FIM evaluations, reconstructing $\mathbf W=\mathbf U_d\mathbf V$, feasibility checks, line searches, and rank-one recovery also contribute to the runtime and need not scale proportionally with $d$.

\end{appendices}

\bibliographystyle{IEEEtran}
\bibliography{Reference}

\end{document}